\documentclass[pre,twocolumn,showpacs]{revtex4}
\usepackage{graphicx}
\usepackage{epsfig}
\usepackage{amssymb,amsmath}
\usepackage{bm}
\def\Q{\mbox{\sffamily\bfseries Q}}
\def\T{\mbox{\sffamily\bfseries T}}

\def\I{\mbox{\sffamily\bfseries I}}

\def\G{\mbox{\sffamily\bfseries G}}

\def\nabbold{\mbox{\boldmath $\nabla$\unboldmath}}
\def\otens{\mbox{\boldmath $\Omega$\unboldmath}}
\def\sigbold{\mbox{\boldmath $\sigma$\unboldmath}}
\def\kapbold{\mbox{\boldmath $\kappa$\unboldmath}}

\def\sigop{\mbox{\boldmath $\sigma$\unboldmath$^{\tiny OP}$}}
\def\n{\mbox{\boldmath $n$\unboldmath}}

\def\ubold{\mbox{\boldmath $u$\unboldmath}}
\def\hatnui{\mbox{\boldmath ${\hat{\nu}}_i$\unboldmath}}

\def\3dots{\:\raisebox{-0.5ex}{$\stackrel{\textstyle.}{:}$}\:}
\def\beq{\begin{equation}}
\def\eeq{\end{equation}}
\def\bea{\begin{eqnarray}}
\def\eea{\end{eqnarray}}
\begin{document}
\title{Routes to spatiotemporal chaos in the rheology of nematogenic fluids
}
\author{Moumita Das$^1$}
\email{moumita@physics.iisc.ernet.in}
\author{Buddhapriya Chakrabarti$^2$}
\email{buddho@physics.umass.edu}
\author{Chandan Dasgupta$^1$}
\author{Sriram Ramaswamy$^1$}
\author{A.K. Sood$^1$}
\email{asood@physics.iisc.ernet.in}
\affiliation{$^1$ Department of Physics, Indian Institute of Science, Bangalore 560012 INDIA\\
$^2$ Department of Physics, University of Massachusetts, Amherst, MA 01003}
\date{\today}
\begin{abstract}
With a view to understanding the ``rheochaos'' observed in recent experiments 
in a variety of orientable fluids,
we study numerically the equations of motion of the spatiotemporal evolution of the
traceless symmetric order parameter of a sheared nematogenic fluid.
In particular we establish, by decisive numerical tests, that the 
irregular oscillatory behavior seen in a region of parameter space where 
the nematic is not stably flow-aligning is in fact spatiotemporal chaos. 
We outline the dynamical phase diagram of the model and study the route to the
chaotic state. We find that spatiotemporal chaos in this system sets in via 
a regime of {\em spatiotemporal intermittency}, with a power-law distribution 
of the widths of laminar regions, as in H. Chat\'{e} and P. Manneville, Phys. Rev. Lett. {\bf 58}, 112 (1987). Further, the evolution of the histogram of band sizes 
shows a growing length-scale as one moves from the chaotic towards the flow aligned phase. 
Finally we suggest possible experiments which can observe the intriguing 
behaviors discussed here. 
\end{abstract}
\pacs{61.30.-v,95.10.Fh,47.50.+d}
\maketitle 
\section{Introduction}
\begin{figure}
\includegraphics[width=9cm,height=15cm]{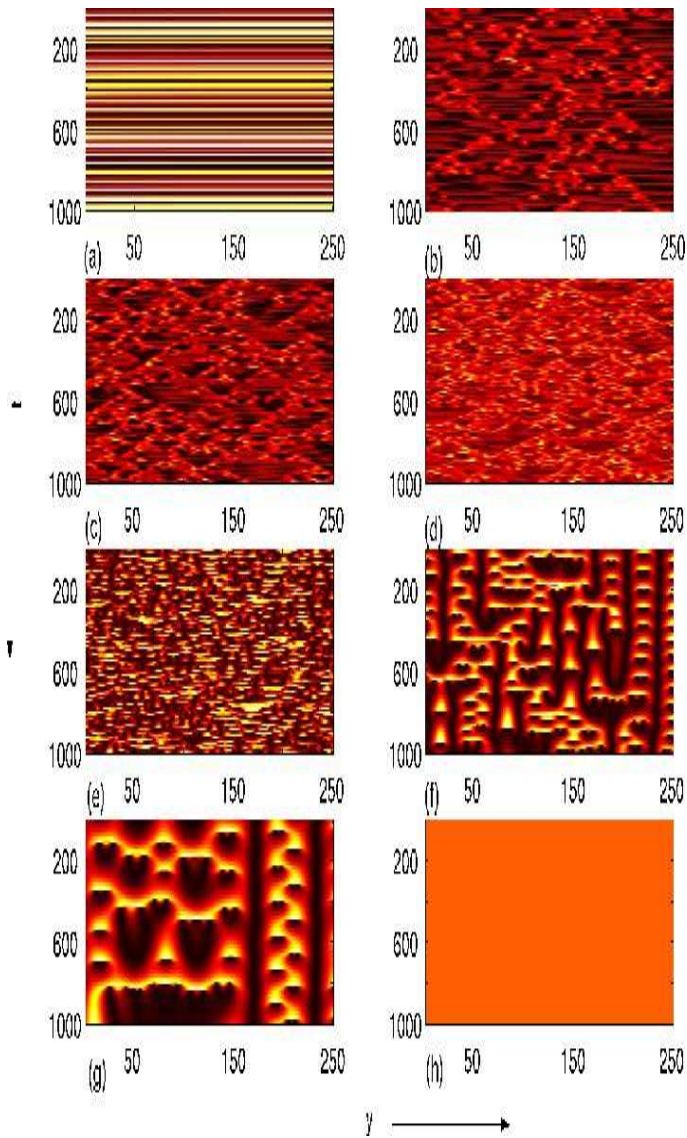}
\caption{\label{approachRC} (Color online) Space-time plots (spatial variation along
abscissa and time along ordinate) of the shear stress for
$\dot{\gamma}=4.0$ and (a) $\lambda_k=$ 1.11 (time-periodic, spatially homogeneous),
(b) to (d), $\lambda_k=$ 1.12,1.13,1.15 (spatiotemporally intermittent), 
(e) $\lambda_k=$ 1.22 (spatiotemporally 
chaotic), (f) and (g), $\lambda_k=$ 1.25 and 1.27 (chaotic to aligning) and 
(h) $\lambda_k=$ 1.28 (aligned) 
(colormap used: black(low shear stress)
$\rightarrow$ red $\rightarrow$ yellow(high shear stress). Slices taken
from system of size $L=5000$.}
\end{figure}

The intriguing rheological behavior of solutions of entangled wormlike micelles 
has been the subject of a large number 
of experimental and theoretical studies in recent 
years~\cite{Larson:99,Israelachvili:85}. These 
long, semiflexible cylindrical objects, whose length distribution 
is not fixed by chemical synthesis and can vary reversibly when 
subjected to changes in temperature, concentration, salinity and 
flow, have  radii $\sim$ 20-25 $\mbox{\AA}$, persistence lengths 
$\sim$ 150 $\mbox{\AA}$ and average lengths upto 
several microns. Like polymers they entangle above a critical 
concentration and show pronounced viscoelastic effects. However, 
unlike covalently bonded polymers, these ``living polymers'' can break and 
recombine reversibly in solutions, with profound consequences 
for stress relaxation and rheology in the form of shear 
banding\cite{Makhloufi:95,Mair:96,Berret:04}, and rheological 
chaos\cite{Bandyopadhyay:99,Bandyopadhyay:00,Cates:02,Fielding:04,Chakrabarti:04,Das:04,
Chakrabarti:03}.
Measurements \cite{Maxwellian,Rehage:91} 
report monoexponential relaxation of the viscoelastic response 
in accordance with the Maxwell model of viscoelasticity.
However, for wormlike micelles of 
CTAT\cite{Bandyopadhyay:00,Bandyopadhyay:00a} at concentration
$1.35$ wt.\%, the fit to the Maxwell model is very poor, and
the Cole-Cole plot deviates from the semi-circular behavior expected
in Maxwellian systems and shows an upturn at high frequencies. 
This deviation from Maxwellian behavior is possibly due to the comparable values 
of timescales associated with reptation ($\tau_{rep}$) and reversible scission
($\tau_{b}$) in this system unlike in other wormlike micellar
systems where the differences in the time scales $\tau_{b} << \tau_{rep}$
lead to a `motional averaging' effect.
Further, in the concentrated regime, when the mesh size of the 
entangled micellar network is shorter than the persistence length of the micelles, 
orientational 
correlations begin to appear \cite{Berret:04}. In fact the nature 
of viscoelastic response and the development of long-range orientational 
order at high concentration play an important role in the non-linear 
rheology of wormlike micelles, in particular in shear banding transition 
and rheochaotic behavior\cite{Chakrabarti:04,Das:04}.

In this paper we explore the dynamical phase diagram of the model studied 
in \cite{Chakrabarti:04,Das:04}, with emphasis on the route to spatiotemporal 
chaos. Our primary finding is summarized in Fig. \ref{approachRC}, which 
shows that this route is characterized by spatiotemporal intermittency.  
Before presenting our results in more detail, we cover some necessary background 
material.  

The application of large stresses and strains on wormlike 
micellar solutions can result in a variety of complex rheological 
behavior. Many dilute solutions of wormlike micelles
exhibit a dramatic shear thickening behavior when sheared above 
a certain threshold rate, often followed by the onset of a flow 
instability\cite{Hu:98,Kadoma:98,Wheeler:98}. Experiments have 
observed shear-banded flow in wormlike micellar solutions with 
formation of bands or slip layers of different microstructures 
having very different rheological 
properties\cite{Spenley:93,Berret:94,Olmsted:97,Larson:92,Larson:99,Mair:96,Becu:04}. The shear banding transition is a transition between a 
homogeneous and an inhomogeneous state of flow, the latter being 
characterized by a separation of the fluid into macroscopic domains 
or bands of high and low shear rates. It is associated with a stress 
plateau (above a certain critical shear rate $\dot\gamma_c$ 
where the shear stress $\sigma$ versus shear rate $\dot\gamma$ curve 
is a plateau) in the nonlinear mechanical response.

More recently, rheological chaos or ``rheochaos'' has been observed   
in experiments studying the nonlinear rheology of dilute entangled 
solutions of wormlike micelles formed by a surfactant 
CTAT\cite{Bandyopadhyay:99,Bandyopadhyay:00,Bandyopadhyay:00a,Bandyopadhyay:01}.
Under controlled shear rate conditions in the plateau regime, the 
shear stress and the first normal stress difference show oscillatory 
and more complicated, irregular time-dependence. Analysis of the 
measured time series shows the existence of a positive Lyapunov exponent 
and a finite non-integer correlation dimension characteristic of 
deterministic chaos.

Occurrence of sustained oscillations often of an irregular nature  
have also been reported in some other experiments on complex fluids 
in shear flow. Roux {\it et al.}\cite{Roux:02,Roux:03} have observed 
sustained oscillations of the viscosity near the non-equilibrium, 
layering transition to the ``onion'' state in a lyotropic lamellar 
system consisting of close compact assembly of soft elastic 
spheres\cite{Roux:02,Roux:03,Wunenburger:01}. It has been conjectured 
that the presence of oscillations in the viscosity is due to structural
changes in the fluid, arising out of a competition between an ordering
mechanism that is driven by stress and a slow textural evolution which
destroys the stress induced ordered state. 
``Elastic Turbulence'' in highly elastic polymer 
solutions\cite{Groisman:00} and ``Director Turbulence'' in nematic liquid 
crystals in shear flow\cite{Cladis:92,Manneville:81} are two other examples
of highly irregular low-Reynolds-number flows in complex fluids. Both these 
phenomena are characterized by temporal fluctuations and spatial 
disorder.
Also worth noting is the observation by Ramamohan 
{\it et al.}\cite{Dasan:02} of rheochaos in numerical studies of 
sheared hard-sphere Stokesian suspensions. 

Many complex fluids have nonlinear rheological 
constitutive equations that cannot sustain a homogeneous steady flow. This 
material instability occurs when the stress vs. strain rate curve is 
non-monotonic in nature, admitting multiple strain rates 
$\dot{\gamma}$ at a common stress $\sigma$. Particularly for shear 
flow, it has been shown \cite{Katz:70} that homogeneous flow is 
linearly unstable in a region where the incremental shear viscosity 
is negative, i.e., $d\sigma/d{\dot{\gamma}} < 0$. The system then 
undergoes a separation into two co-existing macroscopic shear bands 
at different shear rates arranged so as to match the total imposed 
shear gradient. Systems where the dynamic variables 
$\sigma$ or $\dot{\gamma}$ are coupled to microstructural quantities 
may admit many other possibilities -- the flow may never be rendered 
steady in time, or it may become spatially inhomogeneous even erratic 
or both. Fielding and Olmsted study one such scenario\cite{Fielding:04} 
in the context of shear thinning wormlike micelles where the flow 
is coupled to the mean micellar length. 

Significantly, Grosso {\it et al.} \cite{Grosso:01} and Rien\"{a}cker {\it et al.} 
\cite{Rienacker:02,Rienacker:02a},
find temporal rheochaos in the dynamics of the passively advected alignment tensor alone.
They study the well-established equations of hydrodynamics for a nematic order
parameter, with material constants corresponding to a situation where stable flow 
alignment is impossible. They consider only spatially {\em homogeneous} 
states \cite{Hess:81}, i.e., they study a set of ordinary differential equations 
for the independent components of the nematic order parameter, evolving in the presence 
of an imposed plane shear flow. They are thus not in a position to explore the 
implications of the observed chaos for shear-banding. 

Other theoretical approaches aimed at explaining the rheological 
chaotic oscillations in a wormlike micellar fluid include those 
by Cates {\it et al.}\cite{Cates:02}. In the shear thickening 
regime Cates {\it et al.}\cite{Cates:02} propose a simple 
phenomenological model for a fluid with memory 
and an underlying tendency to form shear-banded flows, with only 
one degree of freedom -- the shear stress. Recently Aradian and 
Cates\cite{Aradian:03,Aradian:04} have studied a spatially inhomogeneous 
extension of this model, with spatial variation in the vorticity 
direction. Working at a constant average stress 
$\langle \sigma \rangle$, they observe a rich spatiotemporal 
dynamics, mainly seen in what they call 
``flip flop shear bands'' -- a low and a high unstable shear band separated 
by an interface and periodically flipping into one another. For a 
certain choice of parameters they observe irregular 
time-varying behavior, including spatiotemporal rheochaos. 
As in our work, the key nonlinearities in \cite{Aradian:03,Aradian:04} 
arise from nonlinearities in the constitutive relation, not from 
the inertial nonlinearities familiar from Navier-Stokes turbulence.  
An important result \cite{Aradian:03,Aradian:04} is that they are able to 
find complex flow behavior even when the stress vs. shear-rate curve is monotonic. 
In addition, they find rheochaos even in a few-mode truncation where well-defined 
shear bands cannot arise.

In this paper we study a minimal model to explain the complex 
dynamics of orientable fluids, such as wormlike micelles subjected to shear 
flow. We show that 
the basic mechanism underlying such complex dynamical behavior can 
be understood by analyzing the relaxation equations of the alignment 
tensor of a nematogenic fluid, the underlying idea being that wormlike 
micelles being elongated objects will have, especially when 
overlap is significant,
a strong tendency to align in the presence of shear. 
We study equations of motion of the nematic order parameter 
in the passive advection approximation i.e., ignoring the effect of order parameter
stresses on the flow profile which we take to be plane Couette,
incorporating spatial variation 
of the order parameter. We calculate experimentally relevant quantities, 
e.g, the shear stress and the first normal stress difference, and show that,  
in a region of shear rates, the evolution of the stresses is spatiotemporally chaotic. 
Further, in this region the fluid is not homogeneously sheared but 
shows ``dynamic shear banding'' (banded flow with temporal evolution 
of shear bands). A careful analysis of the space-time plots of the 
shear stress show the presence of a large number of length scales in 
the chaotic region of the phase space of which only a few dominant ones 
are selected as one approaches the boundary of the aligned phase. Finally 
we explore the routes to the spatiotemporally chaotic state. The 
transition from a regular state (either temporally periodic and spatially 
homogeneous, or spatiotemporally periodic) to a spatiotemporally chaotic 
one occurs via a series of spatiotemporally intermittent states. By calculating 
the dynamic structure factor of the shear stress, and the distribution of 
the sizes of laminar domains, we can distinguish this intermittent regime  
from the spatiotemporally chaotic and the regular states occurring 
in this model. Finally, we present a nonequilibrium phase diagram showing 
regions where spatiotemporally regular, intermittent and chaotic phases 
are found. 

The paper is organized as follows. In the next section we introduce the 
model and describe in detail the spatiotemporal chaos that we observe, 
along with the routes to chaos. We then conclude with a summary and discussions 
of our results. Our main results on the spatiotemporal nature of 
rheochaos have appeared in an earlier, shorter article\cite{Chakrabarti:04}. 

\section{Spatiotemporal rheological oscillations and chaotic dynamics 
in a nematogenic fluid:}
\label{spatiotemporal-chaos}

\subsection{Model and Methods}

Traditionally, complex rheological behaviors such as plateau 
in the stress vs shear-rate curve\cite{Berret:04}, 
shear-banding\cite{Makhloufi:95,Mair:96,Berret:04} and 
``spurt''\cite{Spurt:96} have been understood through 
phenomenological models for the dynamics of the stress such as 
the Johnson-Segalman (JS)\cite{Johnson:77,Malkus:90} model, which 
produce non-monotonic constitutive relations. In such equations 
the stress evolves by relaxation or by coupling to the velocity 
gradient. For example in the JS model the non-Newtonian 
part of the shear stress $\sigbold$ evolves according to 
\beq
\label{Johnson-Segalman}
\frac{\partial \sigbold}{\partial t} + \ubold \cdot \nabbold \sigbold + 
\sigbold [ \otens - a \kapbold] + [\otens - a \kapbold]^{T} \sigbold
= 2 \mu \kapbold - {\tau_0}^{-1} \sigbold
\eeq
with a stress relaxation time $\tau_0$, an elastic modulus $\mu$ and
a parameter `$a$' (called the slip parameter) controlling the non-affine
deformation. $\ubold$ is the hydrodynamic velocity field and 
$\otens$ and $\kapbold$ are the antisymmetric and symmetric
parts of the rate-of-deformation tensor. A useful point of view, and one that 
unifies such phenomenological descriptions with dynamical models of 
ordering phenomena in condensed matter physics, is that such 
equations of motion for the stress are not fundamental but are 
derived from the underlying dynamics of an {\it alignment tensor} or 
local nematic order parameter $\Q$. Equations of motion for the latter 
are well-established\cite{Leslie:66&68,Ericksen:60&66,Forster:71,OrsayGroup:71,
MPP:72,Hess:75,Stark:03,deGennes:92,deGP:95} in terms of microscopic mechanics 
(Poisson brackets) and local thermodynamics, and naturally include 
both relaxation and flow-coupling terms of essentially the sort seen, 
e.g., in the JS model. The contribution of the order parameter 
$\Q$ to the stress tensor is also unambiguous within such a 
framework, once the free-energy functional $F[\Q]$ governing 
$\Q$ is specified. This approach is particularly appropriate when the 
system in question contains orientable entities, such as the elongated 
micelles of the experiments of\cite{Bandyopadhyay:99,Bandyopadhyay:00,
Bandyopadhyay:00a,Bandyopadhyay:01}. Thus, not worrying about  
properties specific to a wormlike micelle, e.g., the breakage and recombination 
of individual micelles, one can attempt to understand the properties 
of the wormlike-micelle solution by treating it as an orientable 
fluid and analyzing the equations of motion of the nematic order 
parameter. While properties specific to living polymers might play an important 
role in their rheological behavior, the generality of our order parameter description  
encourages us to think that we have captured an essential ingredient for rheochaos.  
As we shall see, this approach leads 
naturally to terms nonlinear in the stress, absent in the 
usual JS equations of motion, which lead ultimately to the chaos 
with which this paper is concerned. 
Refs. \cite{Aradian:03,Aradian:04} found it necessary to 
modify the Johnson-Segalman equation by including terms nonlinear in 
the stress in order to produce chaos.  

We now discuss the relaxation equation of the alignment tensor 
characterizing the molecular orientation of a nematic liquid crystal in 
shear flow. These equations were derived by various groups 
\cite{Leslie:66&68,Ericksen:60&66,Forster:71,OrsayGroup:71,Hess:75,Stark:03,deGennes:92,deGP:95}, using different formalisms resulting in broadly 
similar though not in all cases identical equations of 
motion. We work with the equations of\cite{Hess:75,Borgmeyer:95}, 
so as to make contact with the recent studies\cite{Rienacker:99} 
of purely temporal chaos in the spatially homogeneous dynamics 
of nematic liquid crystals in flow. These authors have extended 
their analysis\cite{Rienacker:02,Rienacker:02a} to include biaxially 
ordered steady and transient states. Their work has revealed a 
transition from a kayaking-tumbling motion to a chaotic one via a 
sequence of tumbling and wagging states. Both intermittency and 
period doubling routes to chaos have been found.

A nematogenic fluid is comprised of orientable objects, such 
as rods or discs, with the orientation of the $i$th particle 
denoted by the unit vector $\hatnui$. In the nematic 
phase there is an average preferred direction of these molecules, 
which distinguishes it from the isotropic 
phase where there is no such preferred direction. The order 
parameter that measures such apolar anisotropy is the traceless 
symmetric ``alignment tensor'' or nematic order parameter  
\beq
\label{Order-Parameter-definition}
\Q_{\alpha\beta}({\bf r})=\frac{1}{N} \sum_{i=1}^{N} \langle ({\nu_\alpha}^i 
{\nu_\beta}^i -\frac{1}{3} \delta_{\alpha \beta})\rangle \delta({\bf r} - {\bf r}^{i}). 
\eeq
built from the second moment of the orientational distribution function. 
By construction, it is invariant under ${\hatnui} \to - {\hatnui}$ 
and vanishes when the ${\hatnui}$ are isotropically distributed.  

Since nematic fluids possess long range directional order, presence
of spatial inhomogeneities would result in deformations of the director 
field and hence cost elastic energy. In general when the variable 
describing an ordered phase is varying in space, the free energy density 
will have terms quadratic in $\nabbold \Q$. Static mechanical 
equilibrium for $\Q$ corresponds to extremising the 
Landau-de-Gennes free-energy functional 
\bea
\label{Landau-deGennes-F} 
F[\Q] &=& \int \mbox{d}^3x [{A \over 2} \Q:\Q  
- \sqrt{2 \over 3} B (\Q \cdot \Q): \Q  
+ {C \over 4} (\Q : \Q)^2  \nonumber \\&+&
{\Gamma_1 \over 2} \nabbold \Q \3dots \nabbold \Q + 
{\Gamma_2 \over 2} \nabbold \cdot \Q \cdot \nabbold \cdot \Q ], 
\eea
with phenomenological parameters $A$, $B$ and $C$ governing the 
bulk free-energy difference between isotropic and nematic phases, and
$\Gamma_1$ and $\Gamma_2$ related to the Frank elastic constants 
of the nematic phase. A macroscopically oriented nematic with 
axis $\hat \n$ has $\Q=s (\hat \n \hat \n - \I / 3)$ (where 
$\I$ is the unit tensor) which defines the conventional (scalar) 
nematic order parameter $s$. For 
$A$ small enough but positive, F has minima at $s=0$ and at $s=s_0\neq 0$. The 
minimum at $s=s_0$ is lower than the one at $s=0$ when 
$A < A_{*}\equiv 2B^{2}/9C$ which corresponds to the (mean-field) 
isotropic-nematic transition. The functional $F$ plays a key role 
in the dynamics of $\Q$ as well.  
The equation of motion for $\Q$ is  
\bea
\label{Order-Parameter-Eqn-Passive-Advection}
{\partial \Q \over \partial t} 
+ {\bf u}\cdot \nabbold \Q 
&=& \tau^{-1} \G 
            +  (\alpha_0 \kapbold + \alpha_1 \kapbold \cdot \Q)_{\rm{ST}} \nonumber\\&+&
\otens \cdot \Q - \Q \cdot \otens,
\eea
where the subscript ${\rm{ST}}$ denotes symmetrization and trace-removal.
${\bf u}$ is the hydrodynamic velocity field,
$\kapbold \equiv (1/2)[\nabbold {\bf u} + (\nabbold {\bf u})^T]$
and $\otens \equiv (1/2)[\nabbold {\bf u} - (\nabbold {\bf u})^T]$
the shear-rate and vorticity tensors respectively. 
The flow geometry imposed is plane Couette with velocity
$\mathbf{u} = \dot{\gamma} y \hat{x}$ in the $\hat{x}$ direction,
gradient in the $\hat{y}$ direction and vorticity in the $\hat{z}$
direction.  $\tau$ is a bare relaxation
time, $\alpha_0$ and $\alpha_1$ are parameters related to flow alignment, 
originating in molecular shapes.
$\G$,  the molecular field conjugate to $\Q$, is given by

\bea
\label{Molecular-Field-Inhomogenities}
\G \equiv
-(\delta F / \delta \Q)_{\rm{ST}}
&=&  -[A \Q - \sqrt{6} B (\Q \cdot \Q)_{\rm{ST}} + C \Q \Q : \Q ] \nonumber \\
&+& \Gamma_1 \nabla^2 \Q +
\Gamma_2 (\nabbold \nabbold \cdot \Q)_{\rm{ST}}
\eea
to the lowest order in a series expansion in powers of $\nabbold \Q$.

Since $\Q$ is a traceless and symmetric second rank $3\times3$ 
tensor it has five independent components. Accordingly, when the 
equation of motion of the alignment tensor are appropriately scaled, 
it is possible to express it in the following orthonormalized basis
\bea
\Q = \sum_i a_i \T_i, \nonumber\\
\T_0 = \sqrt{3/2}(\hat{\bf z} \hat{\bf z})_{\rm{ST}}, \nonumber\\ 
\T_1 = \sqrt{1/2}(\hat{\bf x} \hat{\bf x} - \hat{\bf y} \hat{\bf y}), \nonumber\\ 
\T_2 = \sqrt{2}(\hat{\bf x} \hat{\bf y})_{\rm{ST}}, \nonumber\\ 
\T_3 = \sqrt{2}(\hat{\bf x} \hat{\bf z})_{\rm{ST}}, \nonumber\\ 
\T_4 = \sqrt{2}(\hat{\bf y} \hat{\bf z})_{\rm{ST}},
\label{Order-Parameter-Basis}
\eea
and study the equations of motion of each of the components 
$a_{k},\,\, k = 0, 1, \ldots, 4$\cite{Rienacker:02,Rienacker:02a} projected 
out. 

It has been observed in the absence of spatial variation that 
depending on the model parameters entering the equations, the order 
parameter equations can have different characteristic 
orbits\cite{Rienacker:02,Rienacker:02a}. Possible in-plane states, 
where as the name suggests, the director is in the plane 
of flow determined by the direction of the flow and its gradient, and 
the order parameter components $a_3$, $a_4$ $= 0$ are ``Tumbling'' 
($T$, in-plane tumbling of the alignment tensor), ``Wagging'' ($W$, 
in-plane wagging) and ``Aligning'' ($A$, in-plane flow alignment) 
states. Out of plane solutions, characterized by non-zero values 
of $a_3$ and $a_4$, observed are ``Kayaking-tumbling'' ($KT$, a periodic 
orbit with the projection of the main director in the shear plane 
describing a tumbling motion), ``Kayaking-wagging'' ($KW$, a periodic 
orbit with the projection of the main director in the shear plane 
describing a wagging motion) and 
finally ``Complex'' ($C$) characterized by complicated motion of the 
alignment tensor. This includes periodic orbits composed of sequences 
of $KT$ and $KW$ motion and chaotic orbits characterized by a positive 
largest Lyapunov exponent.

A solution phase diagram based on the various in-plane 
and out-of-plane states for $A = 0$ and $\alpha_1 = 0$ is 
given in\cite{Rienacker:02a}. It is observed that 
$\alpha_1 \neq 0$ gives similar 
results\cite{Chakrabarti:03}. As control parameters, we use
$\lambda_k \equiv -(2/\sqrt{3})\alpha_0$ related to the
tumbling coefficient in Leslie-Ericksen 
theory\cite{Rienacker:02,Rienacker:02a}, and the shear 
rate $\dot{\gamma}$ to study the phase behavior of this system.

It is observed in experiments that the flow curve 
(shear stress {\it vs} strain rate) of a wormlike micellar 
system in shear flow has a rather large plateau region where 
banded flow is believed to occur, and a study of the dynamics 
of traceless symmetric order parameter 
$\Q$ (Eq.\ref{Order-Parameter-Eqn-Passive-Advection}) for a 
sheared nematogenic system, that allows spatial variation, is 
likely to capture this feature. As we shall see later the shear 
banding observed in such systems is dynamic in nature and is 
an important element in the spatiotemporal rheochaos we observe.

Hereafter we express the equations in the orthonormalised 
basis as in Eq.\ref{Order-Parameter-Basis}. As 
in\cite{Rienacker:02,Rienacker:02a}, we rescale time by the 
linearized relaxation time $\tau/A_{*}$ at the mean-field isotropic-nematic 
transition,  and $\Q$ as well by its magnitude 
at that transition. We have set $\alpha_1=0$ in our analysis 
as it seems to have little effect on the dynamical behavior of the 
system\cite{Chakrabarti:03,Rienacker:02,Rienacker:02a} in the parameter 
range studied. Further, we choose $A = 0$ throughout, to make a correspondence to the 
ODE studies of \cite{Rienacker:02,Rienacker:02a}. This places 
the system well in the nematic phase at zero shear, in fact at 
the limit of metastability of the isotropic phase. Distances are 
non-dimensionalized by the diffusion length constructed out 
of $\Gamma_1$ and $\tau/A_*$. The ratio $\Gamma_2 /\Gamma_1$ 
is therefore a free parameter which we have set to unity in 
our study. 

The resulting equations are then numerically integrated using 
a fourth order Runge-Kutta scheme with a fixed time 
step ($\Delta t = 0.001$). For all the results quoted here a 
symmetrized form of the finite difference scheme involving nearest 
neighbors is used to calculate the gradient terms. Thus 
\begin{eqnarray}
\nabla^2 f_{i} = \frac{f_{i+1} + f_{i-1} - 2 f_{i}}{(\Delta x)^2} \nonumber \\
\nabla f_{i} = \frac{f_{i+1} - f_{i-1}}{2 \Delta x}
\label{Finite-Difference-scheme}
\end{eqnarray}
We have checked that our results are not changed if smaller 
values of $\Delta t$ are used. We have further checked that 
the results do not change if the grid spacing is changed 
(i.e. $\Delta x$ is decreased) and more neighbors to the left 
and right of a particular site in question are used to calculate 
the derivative. This gives us confidence that the results quoted 
here do reflect the behavior of a continuum theory and are not 
artifacts of the numerical procedure used. We use boundary 
conditions with the director being normal to the walls. With 
this, we discard the first $6 \times 10^6$ timesteps to avoid 
any possibly transient behavior. We monitor the time evolution 
of the system for the next $5 \times 10^6$ time steps 
(i.e. $t=5000$), recording configurations after every $10^3$ 
steps. We have carried out the study with system sizes ranging 
from $L=100$ to $L=5000$. 

Further, we calculate the contribution of the alignment tensor 
to the deviatoric stress\cite{Rienacker:02a,Forster:74,Doi:81,Olmsted:90}
$\sigop \propto \alpha_0 \G - \alpha_1(\Q \cdot \G)_{ST}$
where $\G$, defined in (Eq.\ref{Molecular-Field-Inhomogenities}), is 
the nematic molecular field, and the total deviatoric stress is 
$\sigop$ plus the bare viscous stress. Since the latter is a constant 
within the passive advection approximation, we can study the rheology 
by looking at $\sigop$ alone. We are aware of the importance of allowing 
the velocity profile to alter in response to the stresses produced by the
order parameter field, and this is currently under study \cite{Debarshini:04}.

While generating the results for the time-series analysis for the 
LS, we run the simulation till $t=20000$, for 
$L=5000$ sized system, recording data at space points at intervals of 
$l=10$. We monitor the space-time evolution of the shear stress (the 
$xy$ component of the deviatoric stress $\sigop$) (referred to as 
$\Sigma_{xy}$) and the first and second normal stress differences
$\Sigma_{xx} - \Sigma_{yy}$ and $\Sigma_{yy} - \Sigma_{zz}$ 
respectively.

\subsection{Results and Discussion}

\subsubsection{Phase Behavior and Dynamic Shear Banding}

In view of prior work on observation of chaos in the local 
equations of motion of the alignment tensor (spatially homogeneous 
version of Eq.\ref{Order-Parameter-Eqn-Passive-Advection}) by 
Rien\"{a}cker {\it et al.}\cite{Rienacker:02,Rienacker:02a} we 
address the following question: Is the phase diagram (in the 
$\dot\gamma-\lambda_k$ plane) affected by allowing spatial 
variation of the order parameter ? We answer this question in 
affirmative, and show that the `C' region of the phase diagram of 
Rien\"{a}cker {\it et al.}\cite{Rienacker:02,Rienacker:02a} corresponding 
to ``complex'' or chaotic orbits {\em broadens} upon incorporating 
the spatial degrees of freedom. In other words, there exist parameter 
ranges where the spatially homogeneous system is not chaotic, 
but chaos sets in once inhomogeneity is allowed. A result of 
particular interest is the observation of ``spatiotemporally intermittent'' 
(STI) states in a certain range of parameters {\it en route} from 
the regular to the spatiotemporally 
chaotic regimes. Such behavior is by definition not accessible in 
the evolution equations of the spatially homogeneous alignment tensor studied 
in \cite{Rienacker:02,Rienacker:02a}.  
It would be of interest to find chaotic regimes where only two of the five 
independent components of $\Q$ are nonzero. 
Since the number of degrees of freedom per space-point would then be two, such 
chaos would clearly be a consequence of spatial coupling. 
We have not located such a regime so far.

\begin{figure}
\includegraphics[width=8.5cm,height=5.2cm]{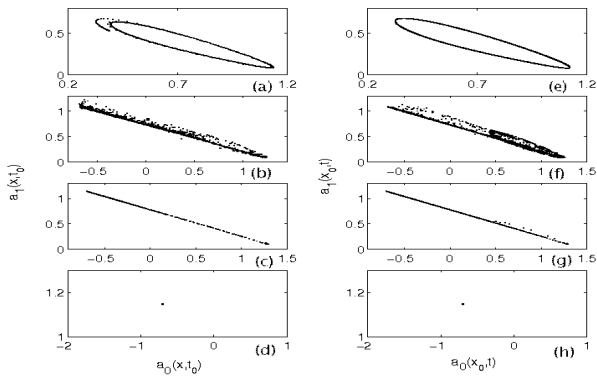}
\caption{\label{Local-Phase-Portraits}Plots showing $a_0(x,t_0)$ 
vs $a_1(x,t_0)$ (left panel) and $a_0(x_0,t)$ vs $a_1(x_0,t)$ (right panel) 
for Periodic [(a),(e)], Chaotic [(b),(f)], (C$\to$A) [(c),(g)] and aligned 
[(d),(h)] regimes.}
\end{figure}

Local phase portraits (orbits obtained when various pairs of order 
parameter components are plotted against each other) illustrate the 
chaotic or orderly nature of the on-site dynamics. Shown in the right 
panels of Fig.\ref{Local-Phase-Portraits} are the local phase portraits 
($a_{1}$ vs $a_{0}$) for a particular point $x_0$ for various values 
of the tumbling parameter $\lambda_{k}$, obtained by holding the 
shear-rate fixed at $\dot{\gamma}=3.5$. We have checked that the 
character of the phase portrait remains 
intact upon going from one space point to another though in the chaotic 
regime there is 
no phase coherence between two such portraits. A closed curve 
corresponding to a limit cycle is seen at $\lambda_k=1.27$, while at 
$\lambda_k=1.3$ (corresponding to the `C' region of the phase space) 
it is space filling. At $\lambda_k=1.35$, as one approaches the region 
where the director aligns with the flow the points reduce to those on 
a line and eventually in the aligning regime ($\lambda_k=1.365$) where 
the director has already aligned with the flow it is represented by 
a point. This assures us that the local dynamics in the spatially 
extended case is similar to that of the ODEs 
of\cite{Rienacker:02,Rienacker:02a}. 

We also construct the spatial analogues of these portraits, i.e., we 
allow the system to evolve till a sufficiently long time (say $t_0$)
and then record the spatial series. Again we get a limit cycle in the 
`T' region of \cite{Rienacker:02,Rienacker:02a},, followed by a space-filling 
curve in the `C' region. As we 
go from the Chaotic towards the Aligning regime, the points arrange 
themselves on a line, and finally in the aligned regime one only 
obtains a point, corresponding to a spatially uniform state. This 
is shown in the left panel for Fig.\ref{Local-Phase-Portraits}.
One should note here that the `T' regime here corresponds to spatiotemporally
periodic states whereas in other regions of parameter space one does
observe states that are temporally periodic but spatially homogeneous (Fig. \ref
{approachRC} (a)) and for such states the local phase portrait corresponding
to spatial variation at a fixed time would be a point and not be a closed curve.

\begin{figure}
\includegraphics[width=8.5cm,height=5.2cm]{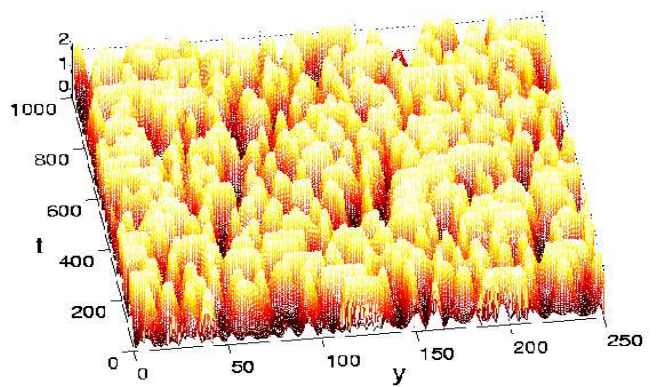}
\caption{\label{Chaotic-Shearstress} (Color online) Space-time behavior of the shear
stress in the chaotic regime, $\dot{\gamma}$=3.678 and $\lambda_k$=1.25.
Slice taken from a system of size $L=5000$.}
\end{figure}

\begin{figure}
\includegraphics[width=8.5cm,height=5.2cm]{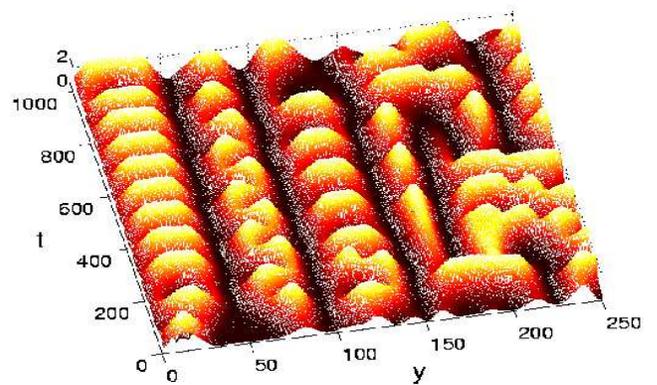}
\caption{\label{Chaotic-Aligning-Shearstress} (Color online) Space-time behavior
(surface plots) of the shear stress in the chaotic to aligning
regime, $\dot{\gamma}=4.05$ and $\lambda_k$=1.25. Slice taken from
a system of size $L=5000$.}
\end{figure}

\begin{figure}
\includegraphics[width=8.5cm,height=5.2cm]{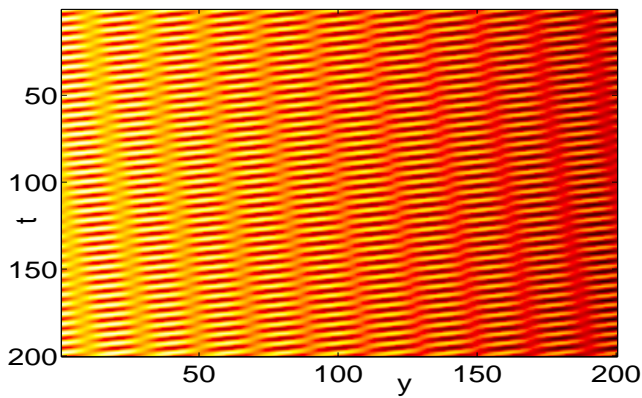}
\caption{\label{spacetimecorSPACETIMEperiodic} (Color online) The 
space-time correlation function in a regime which is both spatially and 
temporally periodic. $\dot\gamma$=3.4,$\lambda$=1.29}
\end{figure}

\begin{figure}
\includegraphics[width=8.5cm,height=5.2cm]{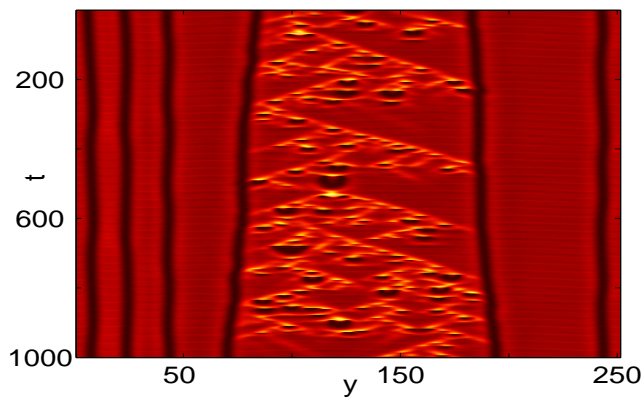}
\caption{\label{coexistSTOandT} (Color online) Space-time evolution of the shear stress
showing co-existence of different dynamical regimes at 
$\dot\gamma$=3.9,$\lambda$=1.12.}
\end{figure}

We now turn to the detailed spatiotemporal structure of the 
phase diagram of this system. We find many interesting phases 
including spatiotemporally chaotic states with a broad distribution 
of length scales (Figs.\ref{approachRC}(e),\ref{Chaotic-Shearstress}), 
 spatiotemporally irregular states in which a few length 
scales are picked up by the system 
(Figs.\ref{approachRC}(f) and (g), \ref{Chaotic-Aligning-Shearstress}), a 
flow aligned phase (Fig. \ref{approachRC}(h)) and also regular states (R)
showing periodicity in both in time and space 
(Fig.\ref{spacetimecorSPACETIMEperiodic}) or that are periodic in time 
and homogeneous in space (Fig.\ref{approachRC}(a)). In addition to these 
states we find the presence of spatiotemporally intermittent states (STI)
(Fig.\ref{approachRC}(b) and (c)). In regions of parameter space we 
have also observed spatially and temporally ordered domains
co-existing with patches characteristic of STI.(Fig.\ref{coexistSTOandT}). 
The parameter values at which these are seen, correspond well 
with those obtained from the phase diagram 
of\cite{Rienacker:02,Rienacker:02a}.

Let us now focus on the parameter region labelled `C' or 
`Complex' in\cite{Rienacker:02,Rienacker:02a}, where we find 
spatiotemporal chaos. This regime is characterized by 
dynamic instability of shear bands as seen in Fig.\ref{Chaotic-Shearstress} 
which shows several distinct events, such as the persistence, 
movement, and abrupt disappearance of shear bands. It is found 
that the typical length scale at which banding occurs is a 
fraction of the system size, though it follows a broad 
distribution. As one moves closer to the phase boundary separating 
the spatiotemporally chaotic state from stable flow alignment, 
the bands become more persistent in time and larger in spatial 
extent as shown in Fig.\ref{Chaotic-Aligning-Shearstress}. 

We have computed the distribution of band sizes or 
spatial ``stress drops'', and looked for the presence 
of dominant length-scales in the system in order to 
obtain a better understanding of the disorderly structure 
of the shear bands as seen in Fig.\ref{Chaotic-Shearstress}, 
and compared it with the behavior seen close to the phase 
boundary (Fig.\ref{Chaotic-Aligning-Shearstress}). Another 
important reason for such an analysis is to rule out any 
hidden periodicity that might be present in the space-time 
profiles of shear stress as shown in Fig.\ref{Chaotic-Shearstress}. 

\begin{figure}
\includegraphics[width=8.5cm,height=10cm]{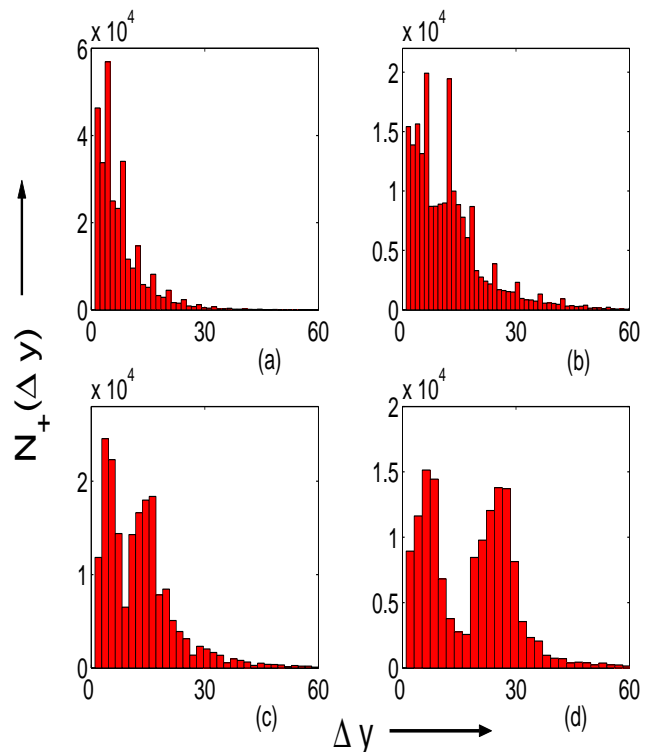}
\caption{\label{Band-distribution} (Color online) Spatial distribution of 
``stress drops'' (corresponding to residence intervals in which 
the shear stress is above a threshold ${\Sigma_0}_{xy}=0.8$) in 
the Chaotic (a) and $C\rightarrow A$ (b,c,d) regimes. $\dot{\gamma}=4.0$ and 
$\lambda_k=$ 1.22 (a),1.24 (b),1.25 (c), 1.27 (d).}
\end{figure}

The ``stress drop'' calculation is outlined below. 
At a given time (say $t_i$), we define a threshold 
${\Sigma_0}_{xy}$, a little above the global mean 
$\langle \Sigma_{xy} \rangle_{y,t}$, and map the 
spatial configuration to a space-time array of 
$\pm 1$: $\tilde{\Sigma_{xy}} = \mbox{sgn}(\Sigma_{xy}-{\Sigma_0}_{xy})$.
Fig.\ref{Band-distribution} shows the histogram of the 
spatial length of intervals corresponding to the $+$state, 
for the Chaotic and the Chaotic to Aligning ($C\rightarrow A$) 
regimes. We have considered configurations extending over 
$L=2500$ spatial points, and the statistics is summed over 
configurations sampled at 5000 times (i.e. $i=1,5000$). As 
expected, the distribution of band lengths in the spatiotemporally 
chaotic regime is fairly broad and roughly exponential in shape, 
whereas as one approaches the Aligning regime, the distribution 
is peaked about a few dominant length-scales. Also, note that as 
one passes from the Chaotic to the Chaotic-to-Aligning (C $\to$ A) state, 
the dominant length-scale associated with the shear bands increases.

\subsubsection{Route to the Spatiotemporally Chaotic State}

\begin{figure}
\includegraphics[width=8.5cm,height=5.2cm]{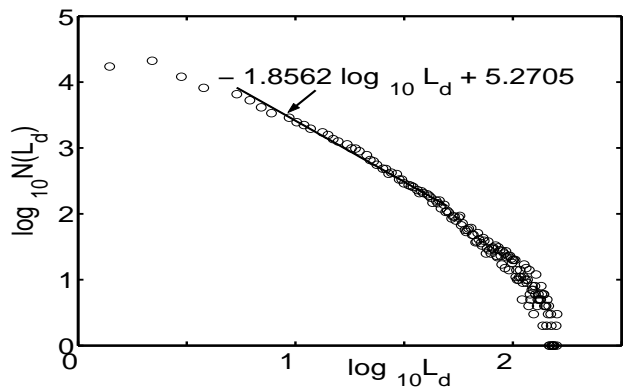}
\caption{\label{laminar-distribution1} Decay of the distribution of laminar domains at the onset of the spatiotemporal intermittent regime (represented by circles) on Log-Log scale. The solid line is a power law fit to the data with an exponent $\psi_{STI}=1.86 \pm 0.05$. }
\end{figure}

\begin{figure}
\includegraphics[width=8.5cm,height=5.2cm]{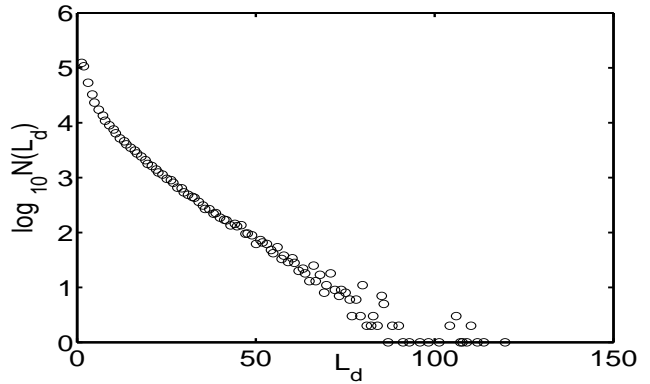}
\caption{\label{laminar-distribution2} Decay of the distribution of laminar domains away from the above-mentioned onset and well within the STI regime (represented by circles) on a semiLog scale.} 
\end{figure}

We now monitor the approach to the spatiotemporally chaotic 
state as a function of the tumbling parameter $\lambda_k$, for 
a fixed value of  $\dot{\gamma}$ ($=3.8$). We observe a sequence 
of states. At low $\lambda_k$ (1.0), the shear stress is periodic 
in time and homogeneous in space Fig.\ref{approachRC}(a). As we 
increase $\lambda_k$, we come across states which are both spatially 
and temporally disordered, (Figs.\ref{approachRC} (b) and (c)) 
consisting of propagating disturbances in a background of highly irregular
local structures, which resemble geometric patterns seen in 
probabilistic cellular 
automata \cite{automata}. The borders 
of the ordered regions evolve like fronts towards each other until 
this region eventually disappears in the chaotic background. These 
states are typical of what is known as 
``spatiotemporal intermittency'' 
(STI)\cite{Chate:88,Kaneko:85,vanHecke:98,Janaki:03}.  
Indeed, it is suggested \cite{vanHecke:98,Chate:87} that 
the transition to fully developed spatiotemporal chaos 
generally occurs via this admixture of complex irregular 
structures (high shear stress) intermittently present with more 
regular low shear regions.  In contrast to 
low dimensional systems where intermittency is restricted to temporal
behavior, STI manifests itself as a sustained regime where 
coherent-regular and disordered-chaotic domains coexist and evolve 
in space and time. Earlier studies of the onset of spatiotemporal 
chaos \cite{Pomeau:86,Chate:88,Kaneko:85,vanHecke:98,JRolf:98,Janaki:03} suggest a relation to 
directed percolation (DP). 
Evidence for DP-like behavior in spatiotemporal
intermittency mostly comes from studies of coupled map lattices 
(CML)\cite{Janaki:03}. Such processes are modeled as a probabilistic 
cellular automaton with two states per site, inactive and active, 
corresponding respectively to the laminar and chaotic domains in 
the case of STI. Studies find that in the STI regime, a laminar 
(inactive) site becomes chaotic (active) at a particular time only 
if at least one of its neighbors was chaotic at an earlier time, 
there being no spontaneous creation of disordered-chaotic sites. Hence 
a disordered site can either relax spontaneously to its laminar state 
or contaminate its neighbors. This feature is analogous to directed 
percolation, and one consequence of this picture is 
the presence of an absorbing state: in STI studies of CMLs,  
once all the sites relax spontaneously to the laminar state, the system 
gets trapped in this state forever, thus the laminar state in STI 
corresponds to an absorbing state in DP. This analogy predicts that STI 
should show critical behavior similar to that associated with DP - power law growth of 
chaotic domains, and characteristic static and spreading exponents. There is
however still no uniformity of opinion on whether spatiotemporal intermittency
belongs to the same universality class as DP as characterised by the critical 
exponents of the DP class. Some studies of coupled map lattices \cite{Chate:88}, 
PDEs \cite{Chate:87,Pang:97} and experiments \cite{Daviaud:90} suggest that, though the 
critical behavior in STI is visually similar to DP, the exponents measured in STI 
are not universal. Other investigations that
have evaluated the exponents at the onset of spatiotemporal intermittency 
in coupled circle map lattices\cite{Janaki:03} and in experimental systems 
\cite{Rupp:02}, claim that this transition indeed falls in the universality 
class of directed percolation.  
Our aim in this context is to macroscopically characterize the
disordered structures in the spatiotemporally intermittent state in terms
of the distribution of the widths of the laminar domains and
study the qualitative connection between 
STI and DP. Accordingly, following 
the analysis in Chat\'{e} {\it et al.} \cite{Chate:88,Chate:87} we study
the decay of the distribution of laminar domains at the onset of intermittency and 
far away from this onset. At each time-step, the spatial series
of the shear stress values are scanned, and the widths of the laminar
regions (regions for which the spatial gradient is less than a sufficiently
small value) are measured and inserted into a histogram. This process is
then cumulated over time, giving the distribution of laminar domains.
Previous studies have found that at the onset of spatiotemporal intermittency, 
this distribution has a power law decay (with a power ranging from 1.5 to 2.0 
in CML studies \cite{Chate:88} and experiments \cite{Daviaud:90} and 3.15 in a variant of
the Swift-Hohenberg eqn. \cite{Chate:87}), while away from the onset, it has an 
exponential decay.
We find a similar behavior. In our work, at a representative point ($\dot{\gamma}=3.9$, 
$\lambda_k=1.116$) at the onset
of intermittency, the distribution of laminar domains has a power law
decay (Fig. \ref{laminar-distribution1}), with an exponent of 1.86 and a standard 
deviation from the data
of $0.05$. Away from this onset ($\dot{\gamma}=4.0$,$\lambda_k=1.13$), 
the decay is close to exponential as evident from Figure \ref{laminar-distribution2}. 


\begin{figure}
\includegraphics[width=8cm]{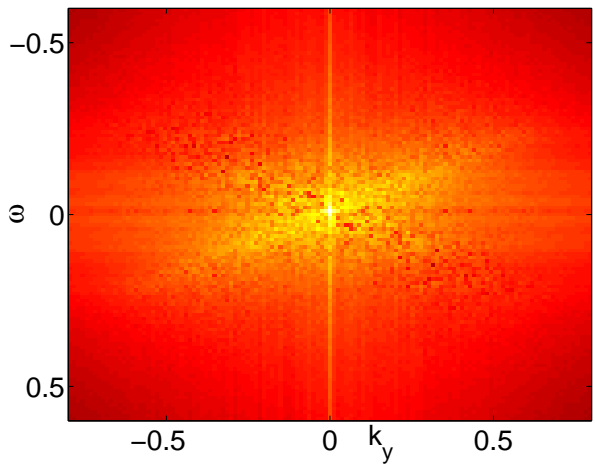}
\caption{\label {Dynamical-Structure-Factor-STI} (Color online)
Pseudocolor plot of the dynamic structure factor 
$S(k_y,\omega)$ in the ``STI'' (spatiotemporally intermittent) regime. Color at any point $\sim$ logarithm (to base 10) of the value of $S(k_y,\omega)$ at that point.
}
\end{figure}

\begin{figure}
\includegraphics[width=8cm]{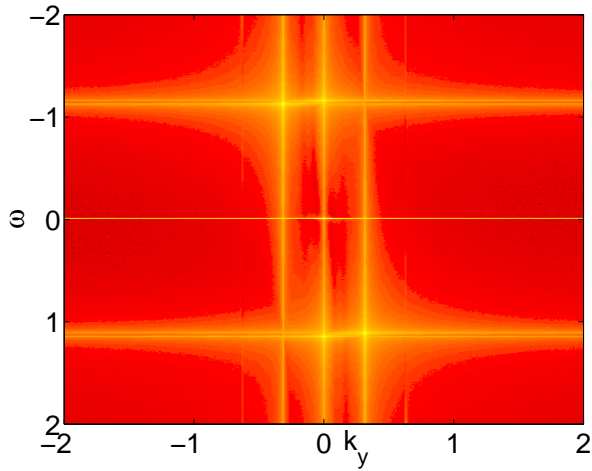}
\caption{\label{Dynamical-Structure-Factor-T} (Color online) Dynamic 
structure factor in the
spatiotemporally periodic regime. The peridicity in time and space is borne out by the straight lines
parallel to the frequency and wave-vector axes respectively.}
\end{figure}

\begin{figure}
\includegraphics[width=8cm]{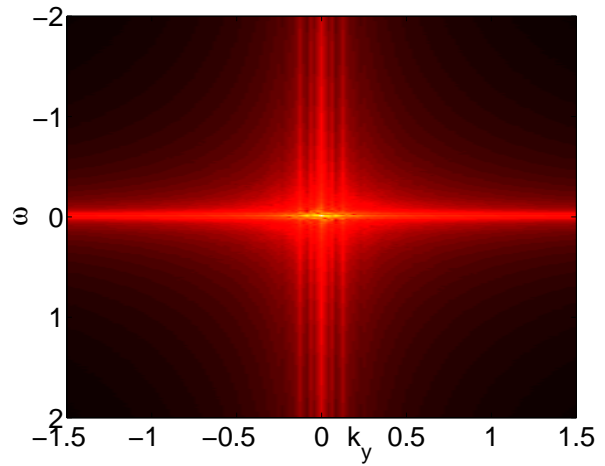}
\caption{\label{Dynamical-Structure-Factor-CtoA} (Color online)
Dynamic structure factor in the
$C \rightarrow A$ regime. The system has almost relaxed to a steady state in time, and spatially there are large 
domains that are flow aligned.}
\end{figure}

Another way of characterizing the time evolution of the coherent and 
disorderly regions in the spatiotemporally intermittent states of  
Figs.\ref{approachRC} (b) and (c) is to  
calculate the dynamical structure factor for the shear stress. 
Fig.\ref{Dynamical-Structure-Factor-STI} shows the dynamic structure 
factor characteristic of this regime. When the system is in the flow 
aligned regime $S(k,\omega)$ has a peak at $k=0$ and $\omega=0$. On 
the other hand in the spatiotemporally intermittent state, the dominant weight
in $S(k, \omega)$ is on lines of $\omega \propto \pm k$, implying disturbances 
with a characteristic speed of propagation. The
front velocity of the cellular automata like patterns seen in the space-time plots 
in this regime can be calculated from the slope of these lines. In 
Figs.\ref{approachRC} (d) and (e), the system is in the chaotic 
regime. As we pass on from the chaotic towards the aligning regime, 
more regular structures are seen to evolve 
(Figs.\ref{approachRC} (f), (g)); the shear bands grow in spatial 
extent and are more long-lived. Fig.\ref{approachRC} (h) shows a 
snapshot of the shear stress in the flow aligned regime.  
Figures \ref{Dynamical-Structure-Factor-T} and \ref{Dynamical-Structure-Factor-CtoA} show the
dynamical structure factor in the spatiotemporally periodic regime and close to the chaotic-aligning phase 
boundary respectively. 

\begin{figure}
\includegraphics[width=8cm]{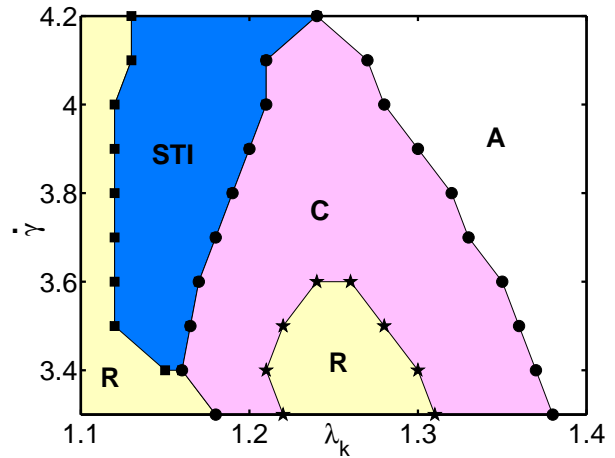}
\caption{\label{phasedigram} (Color online) Phase diagram of the system 
in the $\lambda_k$ vs 
$\dot \gamma$ plane, showing regular, i.e., periodic in time and either 
periodic or homogeneous in space (R), chaotic (C), aligning (A) and spatiotemporally 
intermittent (STI) regimes. Note the reentrant chaotic behavior as a function of $\dot
\gamma$ in a narrow region of $\lambda_k$ and as a function of $\lambda_k$ for
$\dot \gamma < 3.6$.}
\end{figure}

Finally, we present the phase diagram coming out of the previous 
analysis as in the Fig.\ref{phasedigram}.  
The phase diagram was calculated by monitoring the shear stress profile
as well as various components of the nematic order parameter (as
they are strikingly different for the different phases, which is also
 borne out by the difference in the corresponding dynamic structure
 factors)  as a function of the
shear rate $\dot{\gamma}$ and the tumbling parameter $\lambda_k$. 

\subsubsection{Lyapunov structure of the Chaotic State}

Next we try to characterize the chaotic states in our study. In 
studying dynamics of spatio-temporal systems \cite{Pande:00}, one needs to establish 
whether the system is truly in a spatiotemporally chaotic regime 
or can be described by a model with only a few (dominant) independent 
modes. So from the multivariate time-series generated by such 
systems, one tries to compute quantities analogous to the invariant 
measures used to characterize low dimensional chaos. However, true spatiotemporal
chaos corresponds to spatially high dimensional attractors, with dimension 
growing with the system's spatial extent, and the estimation of invariants 
such as the correlation dimension can be quite problematic.  
Indeed we find that the chaos that we observe is 
quite high dimensional (embedding 
dimension\cite{Hegger:99,Kennel:92} $m \ge 10$). A reliable estimate of the 
correlation dimensions can be made only from a data train so long as to  
require prohibitively large computational times to generate. 


An alternative approach is to study the Lyapunov spectrum (LS). 
For a discrete $N$ dimensional dynamical system, there exist $N$ 
Lyapunov exponents corresponding to the rates of expansion and/or contraction 
of nearby orbits in the tangent space in each dimension. The LS 
is then the collection of all the $N$ Lyapunov exponents $\lambda_i, i=1:N$, arranged in 
decreasing order. The LS is very useful in the characterization of a chaotic 
attractor. Useful quantities that can be calculated from the LS are the
number of positive Lyapunov exponents $N_{\lambda_+}$ and sum of the positive Lyapunov exponents
$\sum_{\lambda_{+}}$.
In fact the sum of the positive Lyapunov exponents provides an upper bound for the
so called Kolmogorov-Sinai entropy $h$ which quantifies the mean rate of 
growth of uncertainty in a system subjected to small perturbations. 
In many cases, $h$ is well approximated by the sum $\sum_{\lambda_+}$ \cite{Ruelle:85}. 
Both these quantities have been found to scale extensively with system size in spatiotemporally chaotic systems.
For dynamical systems with only a few effective degrees of freedom, it is straightforward to compute the LS.
However for extended systems with a large number of degrees of freedom, even 
a few hundred,
it runs into severe difficulties because of the inordinately large  
computing time and memory space required. In such situations it is important to make use of techniques that 
derive information about the whole system by analyzing comparatively small systems 
with exactly the same dynamical behavior \cite{Carretero-Gonzalez:99,Orstavik:00,
Carretero-Gonzalez:web}. 
It has been widely observed that the LS for spatiotemporal systems is an extensive measure
\cite {Ruelle:82} and is associated with a  
rescaling property \cite{Carretero-Gonzalez:99,Orstavik:00,Carretero-Gonzalez:web}
i.e., the LS of a subsystem, when suitably rescaled can give rise to the LS of the whole system
\cite{rescaling}.
The volume rescaling property for the LS in spatiotemporally chaotic systems also implies that extensive
(size dependent) quantities such as $\sum_{\lambda_+}$ and $N_{\lambda_+}$ scale
with not only the system size but also the subsystem size.
Hence, instead of trying to study the spectrum and related quantities in a system of large size N, 
one could confine the analysis to relatively small, more manageable 
subsystems of size $N_s$ i.e., at space points $j$ in an interval 
$i_0 < j < i_0 + N_s -1$ (where $i_0$ is an arbitrary reference point), 
and study the scaling of related quantities 
with subsystem size $N_s$\cite{Hegger:99}. Thus, instead of trying to implement the 
correlation-dimension method for our spatially extended problem, we 
study the LS \cite{Hegger:99,Sano:85}. Further, instead 
of studying systems of ever-increasing size, we look at subsystems of 
size $N_s$ in a given large system of size $N$. 

\begin{figure}
\includegraphics[width=8.5cm,height=5.2cm]{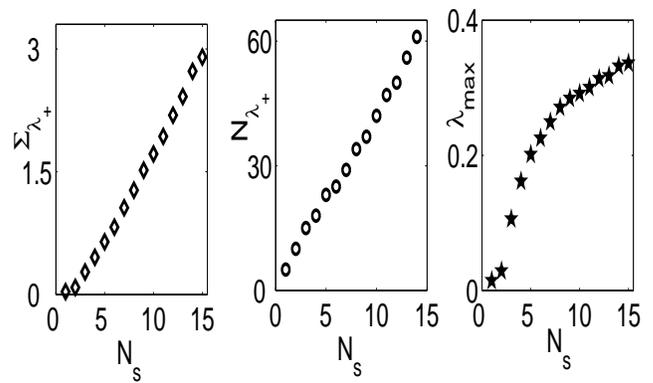}
\caption{\label{Npositivelambda}\small Sum of positive Lyapunov exponents 
(left panel), Number of positive Lyapunov exponents (middle panel) and the 
largest Lyapunov exponent (right panel) as functions of subsystem size 
$N_s$, for $\dot{\gamma}=3.678$, $\lambda_k=1.25$. Embedding dimension 
for the time series of each space point is 10.($i_0$=101, see text).}
\end{figure}

For spatiotemporal chaos we expect to find that the number of positive 
Lyapunov exponents grows systematically with $N_s$. This is seen in 
Fig.\ref{Npositivelambda}. For both figures in Fig.\ref{Npositivelambda}
, we carry out the procedure with two different reference points $i_0$ 
and find essentially the same curves. Furthermore, it has been reported 
in many studies of spatiotemporally chaotic 
systems\cite{Carretero-Gonzalez:99,Orstavik:00,Carretero-Gonzalez:web} 
that when calculating the subsystem LS for increasing
subsystem size $N_s$, one finds that the Lyapunov exponents of two 
consecutive sizes are interleaved, i.e. the $i$th Lyapunov exponent 
$\lambda_i$ for the sub-system of size $N_s$ lies between the $i$th and 
($i+1$)th Lyapunov exponent of the subsystem of size $N_s + 1$. A direct 
consequence of this property is that with increasing subsystem size 
$N_s$, the largest Lyapunov exponent will also increase, asymptotically 
approaching its value corresponding to the case when the subsystem size 
is of the order of the system size. This trend is clearly seen in 
Fig.\ref{Npositivelambda} (b).
 
\subsection{Conclusions}

In summary, we have proposed a mechanism by which one might explain the 
chaotic and irregular rheological response of soft materials in shear 
flow, wormlike micelles in particular. The main idea brought out in this 
paper is that the coupling of orientational degrees of freedom in a complex 
fluid with hydrodynamic flow can lead to spatiotemporal chaos for low 
Reynolds number flows. In particular, we have demonstrated that the 
nonlinear relaxation of the order parameter in nematogenic fluids, together 
with the coupling of nematic order parameter to flow, are key ingredients 
for rheological chaos. 
The broad idea that nonlinearities in the 
stress and spatial inhomogeneity are essential is a feature that our 
work shares with \cite{Fielding:04} and \cite{Aradian:03,Aradian:04}.

We should note here that there could be more than one mechanism at work 
in producing rheochaos. The mechanism observed in our study is that
the system exhibits chaos in its local temporal dynamics,
and then these localized regions mutually interact with one another to generate
spatial disorder. Fielding {\it et al.} \cite{Fielding:04} however have found 
spatiotemporally chaotic
rheological behavior in a model system whose local dynamics does not show chaos,
but incorporating the spatial degrees of freedom makes it chaotic.
Note also that in the parameter range we have studied so far 
the equilibrium phase of the system, in the absence of shear, is 
nematic. We made this choice to facilitate comparison with the work of 
\cite{Rienacker:02,Rienacker:02a}; a better choice from the point of 
view of the experiments on wormlike micelles would be to work in 
the isotropic phase, with a substantial susceptibility to nematic ordering. 
We do not know if shear produces chaos in that situation, although it seems 
likely. It is also worth investigating whether nematics with stable 
flow-alignment at low shear-rates can go chaotic at higher rates of flow. 
That the structures in the transitional region between order and 
spatiotemporal chaos are similar to those in directed percolation, 
as in some other systems undergoing the transition to spatiotemporal 
chaos, is interesting and suggests a direction for possible experimental 
tests. 
 
We now comment on experiments which can test some of the ideas proposed in this paper. 
The dynamics of the alignment tensor can be studied in rheo-optical 
experiments on dichroism\cite{Mewis:97}, flow birefringence and rheo-small angle 
light scattering\cite{Berghausen:97}. Flow birefringence experiments carried 
out in the last decade have shed light on shear banding and orientational properties 
of micellar solutions \cite{Berret:04}. Small angle neutron scattering experiments, using
a two-dimensional detector, have also been used to analyse the orientational degrees 
of a micellar fluid in shear flow: the presence and proportions of the isotropic and nematic 
phases under shear, as well as the order parameter of the shear induced nematic phase in 
such systems have been studied \cite{Berret:04}. In order to investigate rheochaotic
behavior in space and time in systems such as under consideration, one could use these
rheo-optical techniques and try to look for the irregularities in 
the spatial distribution of band sizes and their temporal persistence, in a regime in
the nematic phase where the micelles are not flow aligned. 
Further, very recently, spatio-temporal dynamics of 
wormlike micelles in shear flow has been studied using high-frequency ultrasonic 
velocimetry\cite{Becu:04}, and various dynamical regimes including slow 
nucleation and growth of a high-shear band and fast oscillations of the band 
position have been observed, though the complex fast behavior reported is not 
chaotic. 

We thank G. Ananthakrishna and R. Pandit for very useful discussions,
and SERC, IISc for computational facilities. MD acknowledges support
from CSIR, India, and CD and SR from DST, India through the Centre for
Condensed Matter Theory.


\begin{thebibliography}{}
\bibitem{rescaling} Such a rescaling approach typically consists of studying the evolution of a relatively small
$N_s$ dimensional subsystem (which has the same equations of motion as the original, large,
$N$ dimensional system), and approximating the LS of the latter by rescaling the LS of the
former by the ratio of the volumes $N/N_s$.
\end{thebibliography}
\end{document}